\documentclass[11pt] {article}      

\usepackage{amsmath, amssymb, amsthm, eucal}  
\usepackage{latexsym, graphicx}
\usepackage{times}
\usepackage[T1]{fontenc}     
\usepackage{float}                 
\usepackage{tikz}
\usetikzlibrary{matrix, arrows}
\usepackage{authblk}
\def\smalll{\scriptsize}
\numberwithin{equation}{section}
\numberwithin{figure}{section}
\begin{document}

\title{\sf Coupled-Ring Resonance and Unitary Groups}

\author[1]{Jerzy Kocik}
\author[2]{Mohammad Sayeh}          
\affil[1]{\smalll Department of Mathematics, Southern Illinois University, Carbondale, IL 62901, email: jkocik@siu.edu}
\affil[2]{\smalll Department of Electric Engineering, Southern Illinois University, Carbondale, IL 62901, email: sayeh@siu.edu}
\date{}

\maketitle

\begin{abstract}
The group-theoretic content of photonic coupled microrings resonance phenomena is shown,
in particular,  an interesting emergence of pseudo-unitary group.
The application to the resonance condition in a tri-microring configuration is solved exactly.  
A practical application of this work will be in the resonance frequency tuning based on the coupling coefficient, in particular, 
the Mach-Zehnder interferometer approach was analyzed for the coupling modulation.
\\[5pt]
\\[3pt]
\end{abstract}


\def\xxx{blue}

\section{Introduction}

An optical resonator, passive or active, is one of the important building block in any precision optically-based system.  
Applications range from lasers, filters, and modulators to gyroscopes and biosensors \cite{m1,m2, m3, Rabus, Yariv,Agrawal}.  
Coupled resonators are being used in enhancing the resonators performance by mode selections and narrow-band transmission/reflection \cite{Vahala}. 
The coupling modulation was shown to improve the output characteristics of the microring resonator \cite{sacher1,sacher2}.  
In this work, we introduce yet another resonance property improvement via coupling mechanism, namely, coupling factor frequency tuning.  The tuning of the resonance frequency via the change of optical pathlength suffers from interfering with the resonator free spectral range (FSR).  The coupling modulation presented here is free of this limitation. The analysis of the tri-microring configuration is carried out by using the method of the group-theoretic properties and eigenvector analysis.   
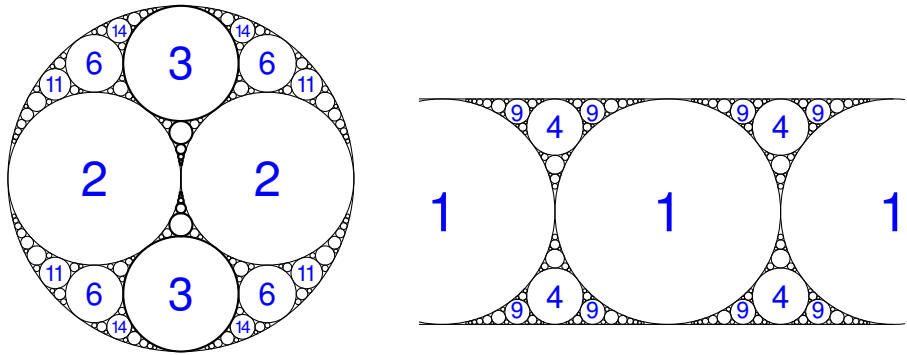
\begin{figure}[H]
\centering
\begin{tikzpicture}[scale=2.3]

\draw (0,0) circle (1);

\foreach \a/\b/\c   in {
1 / 0 / 2 
}
\draw (\a/\c,\b/\c) circle (1/\c)
          (-\a/\c,\b/\c) circle (1/\c);

\foreach \a/\b/\c in {
0 / 2 / 3 ,
0 /4 /15 ,
0 / 6 / 35 , 
0 / 8/ 63,
0 /10 / 99,
0 / 12 / 143
}
\draw[thick] (\a/\c,\b/\c) circle (1/\c)
          (\a/\c,-\b/\c) circle (1/\c) ;

\foreach \a/\b/\c/\d in {
3 / 4 /6, 	8 / 6 / 11,	5 / 12/ 14,	15/ 8 / 18,	8 / 12 / 23,	7 / 24 / 26,
24/	10/	27, 	21/	20/	30, 	16/	30/	35, 	3/	12/	38, 	35/	12/	38, 	24/	20/	39, 	9/	40/	42,
16/	36/	47, 	15/	24/	50, 	 48/	14/	51, 	45/	28/	54, 	24/	30/	59, 	40/	42/	59, 	11/	60/	62,
21/	36/	62, 	48/	28/	63, 	33/	56/	66, 	63/	16/	66, 	8/	24/	71, 	55/	48/	74, 	24/	70/	75,
48/	42/	83, 	80/	18/	83, 	13/	84/	86, 	77/	36/	86, 	24/	76/	87, 	24/	40/	87, 	39/	80/	90
}
\draw (\a/\c,\b/\c) circle (1/\c)       (-\a/\c,\b/\c) circle (1/\c)
          (\a/\c,-\b/\c) circle (1/\c)       (-\a/\c,-\b/\c) circle (1/\c) 
;
\node at (-1/2,0) [scale=1.7, color=\xxx] {\sf 2};
\node at (1/2,0) [scale=1.7, color=\xxx] {\sf 2};
\node at (0,2/3) [scale=1.6, color=\xxx] {\sf 3};
\node at (0,-2/3) [scale=1.6, color=\xxx] {\sf 3};
\node at (1/2,2/3) [scale=1.1, color=\xxx] {\sf 6};
\node at (-1/2,2/3) [scale=1.1, color=\xxx] {\sf 6};
\node at (1/2,-2/3) [scale=1.1, color=\xxx] {\sf 6};
\node at (-1/2,-2/3) [scale=1.1, color=\xxx] {\sf 6};
\node at (8/11,6/11) [scale=.77, color=\xxx] {\sf 1$\!$1};
\node at (-8/11,6/11) [scale=.77, color=\xxx] {\sf 1\!1};
\node at (8/11,-6/11) [scale=.77, color=\xxx] {\sf 1\!1};
\node at (-8/11,-6/11) [scale=.77, color=\xxx] {\sf 1\!1};
\node at (5/14,6/7) [scale=.6, color=\xxx] {\sf 1\!4};
\node at (-5/14,6/7) [scale=.6, color=\xxx] {\sf 1\!4};
\node at (5/14,-6/7) [scale=.6, color=\xxx] {\sf 1\!4};
\node at (-5/14,-6/7) [scale=.6, color=\xxx] {\sf 1\!4};
\end{tikzpicture}
\qquad
%
%
\begin{tikzpicture}[scale=1.5, rotate=90, shift={(0,2cm)}]  
\clip (-1.25,-2.1) rectangle (1.1,2.2);
\draw (1,-2) -- (1,3);
\draw (-1,-2) -- (-1,3);
\draw (0,0) circle (1);
\draw (0,2) circle (1);
\draw (0,-2) circle (1);

\foreach \a/\b/\c in {
3/4/4,  5/12/12,  7/24/24, 9/40/40  
}
\draw (\a/\c,\b/\c) circle (1/\c)    (\a/\c,-\b/\c) circle (1/\c)
          (-\a/\c,\b/\c) circle (1/\c)    (-\a/\c,-\b/\c) circle (1/\c)   ;

\foreach \a/\b/\c in {
8/    6/   9, 	
15/   8/   16 , 	
24/  20/  25, 	
24/  10/  25, 	
21/  20/  28,
16/	30/	33,    
35/	12/	36,
48/	42/	49,
48/	28/	49,
48/	14/	49,
45/	28/	52,
40/	42/	57,
33/	56/	64,
63/	48/	64,
63/	16/	64,
55/	48/	72,
24/	70/	73,
69/	60/	76,
80/	72/	81,
64/	60/	81,
80/	36/	81,
80/	18/	81
}
\draw (\a/\c, \b/\c) circle (1/\c)          (-\a/\c, \b/\c) circle (1/\c)
          (\a/\c,-\b/\c) circle (1/\c)         (-\a/\c,-\b/\c) circle (1/\c)
          (\a/\c,2-\b/\c) circle (1/\c)       (-\a/\c,2-\b/\c) circle (1/\c)
          (\a/\c,2+\b/\c) circle (1/\c)       (-\a/\c,2+\b/\c) circle (1/\c)
          (\a/\c,-2+\b/\c) circle (1/\c)       (-\a/\c,-2+\b/\c) circle (1/\c)
;
\node at (0,0) [scale=1.9, color=\xxx] {\sf 1};
\node at (0,-2) [scale=1.9, color=\xxx] {\sf 1};
\node at (0,2) [scale=1.9, color=\xxx] {\sf 1};
\node at (-3/4,1) [scale=1.1, color=\xxx] {\sf 4};
\node at (-3/4,-1) [scale=1.1, color=\xxx] {\sf 4};
\node at (3/4,1) [scale=1.1, color=\xxx] {\sf 4};
\node at (3/4,-1) [scale=1.1, color=\xxx] {\sf 4};
\node at (7/8,8/6) [scale=.8, color=\xxx] {\sf 9};
\node at (7/8,-8/6) [scale=.8, color=\xxx] {\sf 9};
\node at (7/8,4/6) [scale=.8, color=\xxx] {\sf 9};
\node at (7/8,-4/6) [scale=.8, color=\xxx] {\sf 9};
\node at (-7/8,8/6) [scale=.8, color=\xxx] {\sf 9};
\node at (-7/8,-8/6) [scale=.8, color=\xxx] {\sf 9};
\node at (-7/8,4/6) [scale=.8, color=\xxx] {\sf 9};
\node at (-7/8,-4/6) [scale=.8, color=\xxx] {\sf 9};
\end{tikzpicture}
\caption{Apollonian Window (left) and Apollonian Belt (right)}
\label{fig:Apollo}
\end{figure}

The present investigation is motivated by a greater problem.  
The so-called Apollonian integral disk packings
are fractal arrangements of disks / circles, the curvatures of which are all integral, see Fig. \ref{fig:Apollo} for two examples 
(the numbers inside the circles represent their curvatures).
Such integral Apollonian disk packings are classified and some number-theoretical properties are found \cite {jk,LMW,N}. 
Since the ratios of the circumferences are all rational, fragments of such packings might display 
interesting resonance properties with possible applications in photonic systems.

The interest in these structures is increasing due to its many connections 
to various phenomena known in physics,
like the Hofstadter butterfly 
representing the energy levels of quantum Hall conductivity \cite{IS} ,
or spin networks \cite{jk3} 
-- one of the tools of the quantum gravity project, to mention a few.
This paper investigates an example directed at such application problems.

\section{Basics}
\label{sec:basics}

There have been many efforts to study and understand optical resonators since the advent of lasers.  
It is crucial to have a stable and single-mode operation of the laser frequency in precision interferometric applications 
such as displacement and wavelength measurements.  
One of the early and successful resonator was (still is) the Fabry-Perot resonator consists of only two highly reflecting mirrors. 
This resonator has been used in optical spectrum analyzers and optical filters.  
The frequency of resonance is traditionally tuned by varying the optical round-trip path length.  
Also by cascading the resonators, much higher finesse resonance has been achieved \cite{Poon1,Schwelb,Vahala}.  

We analyze a system of three mutually coupled resonators 
where the resonance frequencies can be tuned by adjustment of the coupling factor.  
The resonator system is easily fabricated by a single lithography step \cite{Coldren}.  
Yariv's group have worked on the cascade of coupled resonators to enhance the resonance frequency linewidth \cite{Poon,Poon1}.  
In this work we develop an alternative approach focusing on the closed-form formula relating the frequency to the coupling factor.
\\

\begin{figure}[ht]
\centering
\includegraphics[scale=.7]{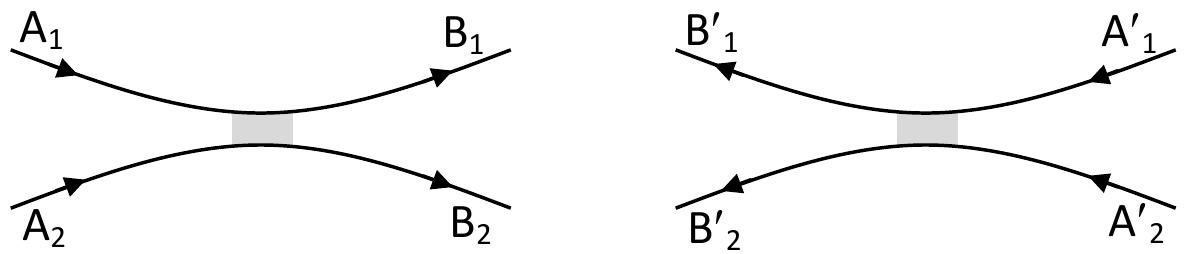}
\caption{Transfer of signal between waveguides}
\label{fig:markov}
\end{figure}

Two optical waveguides, labeled as 1 and 2, are coupled as shown in Figure \ref{fig:markov}. 
The transfer of signals between them is ruled 
by the following transformation \cite{Yariv1,Yariv}
\begin{equation}
\label{eq:trans}   
\left[\begin{array}{r}
                             B_1\cr
                             B_2 \cr \end{array}\right]
=
\left[\begin{array}{rr}
                             \alpha\phantom{^*} &  \beta\phantom{^*} \cr
                             -\beta^* &  \alpha^* \cr \end{array}\right]
\left[\begin{array}{r}
                             A_1\cr
                             A_2 \cr \end{array}\right]   \,,
\end{equation}
where $*$ indicates complex conjugation on a scalar; and $\alpha$ and $\beta$ are the transmission and coupling coefficients, respectively.  
$A_i$ and $B_i$ are the in and out signals, respectively,  in the $i$-th fiber
(see Fig. \ref{fig:markov}, left). 
\\

\noindent
{\bf Observation 1:}  The transition matrix
\begin{equation}
\label{eq:inout}
M = \left[\begin{array}{rrr}
                             \alpha\phantom{^*} &  \beta\phantom{^*} \cr
                             -\beta^* &  \alpha^* \cr \end{array}\right]
\end{equation}
is an element of the special unitary group ${\rm SU}(2)$.  In particular,
$$
M^*M= I
\qquad\hbox{and}\qquad
\det M = |\alpha|^2 + |\beta|^2 = 1    \, ,
$$
where $M^*$ denotes Hermitian conjugation, that is $M^* = \bar M^T$  \ 
(transposition and complex conjugation of the entries).
\\
\\
{\bf Trick:}  Equation \eqref{eq:trans} represents the ``in-out'' transformation.
We may rearrange it to represent a ``waveguide-to-waveguide'' transformation:
\begin{equation}
\label{eq:transx} 
\left[\begin{array}{r}
                             A_2\cr
                             B_2 \cr \end{array}\right]
\ = \ 
\frac{1}{\beta}
\left[\begin{array}{cc}
                             -\alpha &1   \\
                             -1 & \alpha^*      \end{array}\right]
\left[\begin{array}{r}
                             A_1\cr
                             B_1 \cr \end{array}\right]
                             \,.
\end{equation}

\noindent
{\bf Observation 2:}
The waveguide-to-waveguide transition matrix
\begin{equation}
\label{eq:transN} 
N = 
\frac{1}{\beta}
\left[\begin{array}{cc}
                             -\alpha &1   \\
                             -1 & \alpha^*      \end{array}\right]
\end{equation}
satisfies:
$$
N^*\,G\,N \ = \ -G\,,
\qquad
|\det N | = 1\, ,
$$
where $G$ is the inner product form
$$
G = \left[\begin{array}{cr}
                             1  & 0   \\
                             0  &  -1        \end{array}\right]  \,.
$$
In particular, the determinant is:
$$
\det N = \frac{-\alpha\alpha^*+1}{\beta^2}  
= \frac{\beta^*}{\beta} \qquad  \in S^1 \subset \mathbb C\, ,
$$  
where $S^1=\{e^{i\theta}\}$ denotes the unit circle in the complex plane.

~\\
In reference to Fig. \ref{fig:markov}, (right), the law for the case of  reversed direction of signal, right to left,
is ruled by the transposed matrix:
\begin{equation}
\label{eq:transs}  
\left[\begin{array}{r}
                             B'_1\cr
                             B'_2 \cr \end{array}\right]
=
\left[\begin{array}{cr}
                             \alpha &  -\beta^* \cr
                             \beta &  \alpha^* \cr \end{array}\right]
\left[\begin{array}{r}
                             A'_1\cr
                             A'_2 \cr \end{array}\right]\, ,
\end{equation}
where $A$ and $B$  denote the input and output signals, respectively.
The corresponding waveguide-to-waveguide version of such transition becomes
\begin{equation}
\label{eq:transxx}
\left[\begin{array}{r}
                             A'_2\cr
                             B'_2 \cr \end{array}\right]
\ = \ 
\frac{1}{\beta^*}
\left[\begin{array}{cc}
                             \alpha  &-1   \\
                             1 & -\alpha^*      \end{array}\right]
\left[\begin{array}{r}
                             A'_1\cr
                             B'_1 \cr \end{array}\right]\,.
\end{equation} 

~\\
{\bf Observation 3:}  
Propagation along a waveguide contributes phase change by some
$\sigma\in \mathbb C$,  \, $|\sigma|^2 = 1$,
e.g., signal $X$ becomes $X'$ according to
$$
X' = \sigma\,X 
\qquad \hbox{and}\qquad
X= \sigma^* X'   \,.
$$
For two waveguides, each of different length and not in contact, 
we may use a diagonal matrix:
\begin{equation}
\label{eq:phase0} 
\left[\begin{array}{r}
                             A_2\cr
                             B_2 \cr \end{array}\right]
\ = \ 
\left[\begin{array}{cc}
                             \sigma &  0 \cr
                             0 &  \lambda \cr \end{array}\right]
\left[\begin{array}{r}
                             A_1\cr
                             B_1 \cr \end{array}\right]
\,,  \qquad |\sigma|^2 = |\lambda|^2 = 1 \, ,
\end{equation}
where $\lambda$ ($\sigma$)  is phase change along the long external 
(short inner) part of each ring.

Thus, in summary, we have an interesting appearance of the unitary groups:  
${\rm SU}(2)$ for the in-out transformation,
and ${\rm U}(1,1)$ for the waveguide-to-waveguide transformation.
Appendix A gives more details on these groups.

\section{Tri-microring structure}
\label{sec:3circles}

\begin{figure}[ht]
\centering
\includegraphics[scale=.56]{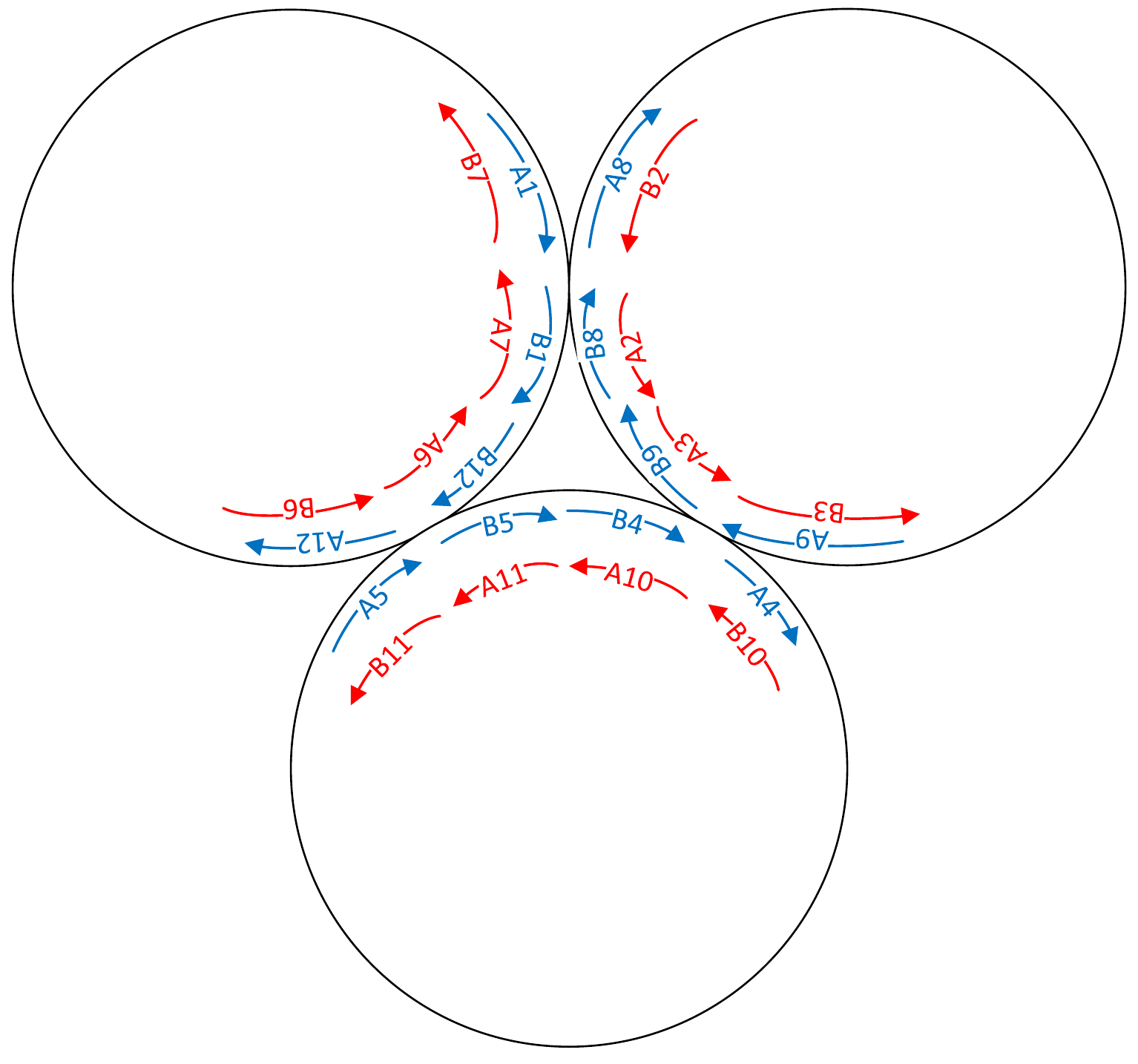}   
\caption{Bidirectional tri-microring configuration}
\label{fig:3circle}
\end{figure}

\begin{figure}[ht]
\centering
\includegraphics[scale=.67]{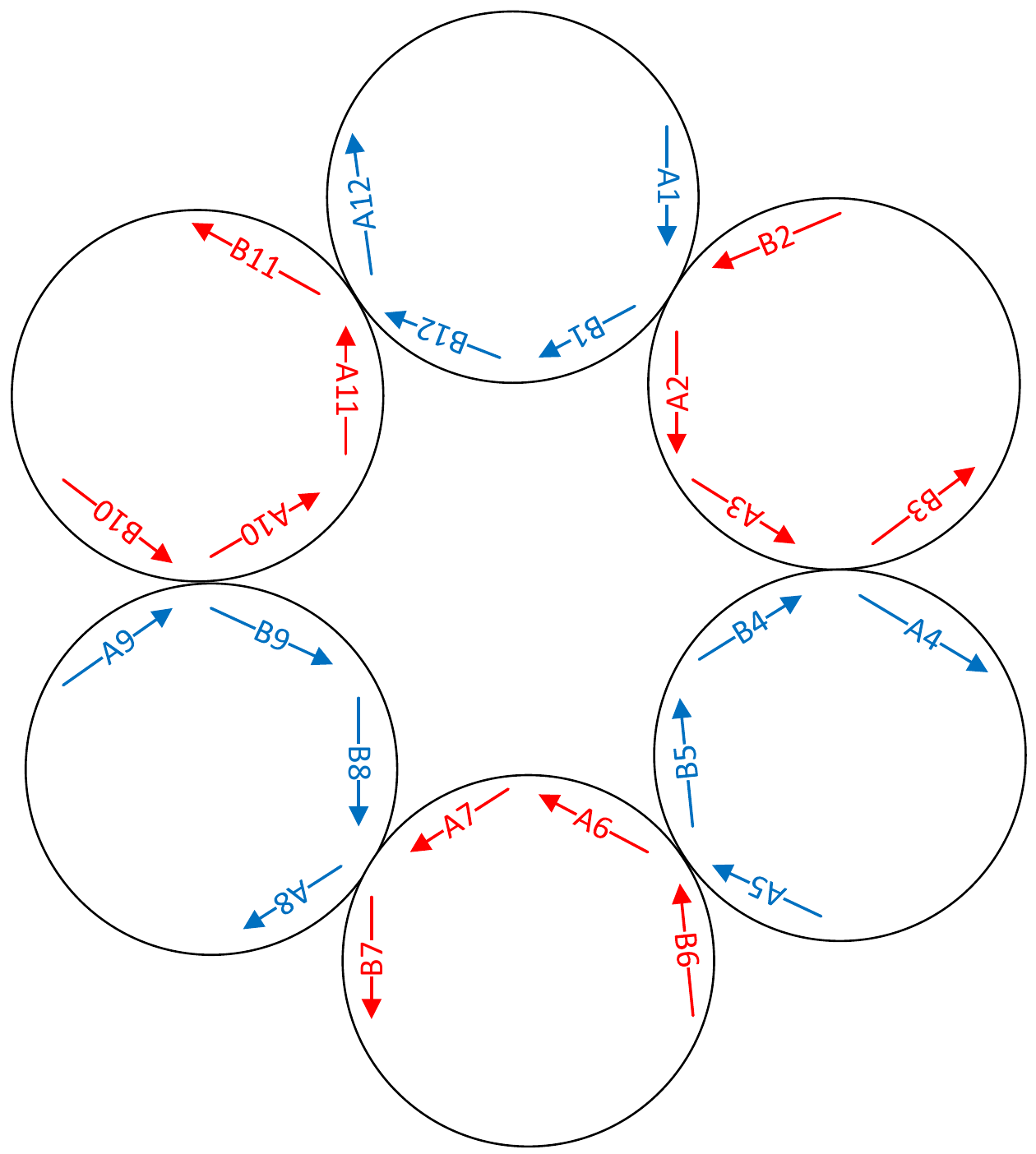}   
\caption{An equivalent unidirectional representation of tri-microring arrangement}
\label{fig:6circle}
\end{figure}

Below, the focus is on the three mutually coupled rings as shown in Fig. \ref{fig:3circle}. 
For the sake of simplicity, a waveguide coupled to the tri-microring arrangement, proving the input/output ports, is not considered in this paper.  
All rings are bi-directional with their corresponding fields as shown.  
In order to utilize the fiber-to-fiber matrices for the resonator fields, 
the three bi-directional ring resonators are converted equivalently 
to the six unidirectional rings as shown in Fig. \ref{fig:6circle}.

%

The chain of transitions runs as follows:
$$
\begin{tikzpicture}[baseline=-0.8ex]
    \matrix (m) [ matrix of math nodes,
                         row sep=1em,
                         column sep=4em,
                         text height=3.8ex, text depth=3ex] 
{
    \left[\begin{array}{r}
                             A_1\cr
                             B_1 \cr \end{array}\right]  \    
& \  
\left[\begin{array}{r}
                             B_2\cr
                             A_2 \cr \end{array}\right]  \   
& \  
\left[\begin{array}{r}
                             B_3\cr
                             A_3 \cr \end{array}\right]
& \  
\left[\begin{array}{r}
                             A_3\cr
                             B_3 \cr \end{array}\right] \to \\
\phantom{ \left[\begin{array}{r}
                             A_3\cr
                             B_3 \cr \end{array}\right] }  \    
& \  
\left[\begin{array}{r}
                             B_4\cr
                             A_4 \cr \end{array}\right]  \   
& \  
\left[\begin{array}{r}
                             B_5\cr
                             A_5 \cr \end{array}\right] \ 
& \  
\left[\begin{array}{r}
                             A_5\cr
                             B_5 \cr \end{array}\right] \phantom{\to} \\
    };
    \path[->]
        (m-1-1) edge node[above] {\smalll $\frac{1}{\beta^*} \begin{bmatrix} \alpha &\!\!\! -1\\  1 &\!\!\! -\alpha^* \end{bmatrix}$} 
                            node[below] {\smalll $N_L$}(m-1-2)
        (m-1-2) edge node[above] {\smalll $\begin{bmatrix} \lambda^* &\!\!\! 0 \\ 0 &\!\!\! \sigma \end{bmatrix}$} (m-1-3)
        (m-1-3) edge node[above] {\smalll $\begin{bmatrix} 0 &\!\!\! 1 \\ 1 &\!\!\! 0 \end{bmatrix}$} (m-1-4)

       (m-2-1) edge node[above] {\smalll $\frac{1}{\beta} \begin{bmatrix} -\alpha &\!\!\! 1\\  -1 &\!\!\! \alpha^* \end{bmatrix}$} 
                            node[below] {\smalll $N'_R$} (m-2-2)
        (m-2-2) edge node[above] {\smalll $\begin{bmatrix} \sigma^* &\!\!\! 0 \\ 0 &\!\!\! \lambda \end{bmatrix}$} (m-2-3)
        (m-2-3) edge node[above] {\smalll $\begin{bmatrix} 0 &\!\!\! 1 \\ 1 &\!\!\! 0 \end{bmatrix}$} (m-2-4)
;
\end{tikzpicture}   
$$
Multiplying the matrices of the string of transitions from node 1 to 5,
{\small
\begin{equation}
\label{eq:1to5}
\begin{bmatrix} 0 &\!\!\! 1 \\ 1 &\!\!\! 0 \end{bmatrix}
\begin{bmatrix} \sigma^* &\!\!\! 0 \\ 0 &\!\!\! \lambda \end{bmatrix}
\frac{1}{\beta} \begin{bmatrix} -\alpha &\!\!\! 1\\  -1 &\!\!\! \alpha^* \end{bmatrix}
\begin{bmatrix} 0 &\!\!\! 1 \\ 1 &\!\!\! 0 \end{bmatrix}
\begin{bmatrix} \lambda^* &\!\!\! 0 \\ 0 &\!\!\! \sigma \end{bmatrix}
\frac{1}{\beta^*} \begin{bmatrix} \alpha &\!\!\! -1\\  1 &\!\!\! -\alpha* \end{bmatrix}
%
\end{equation}
we arrive at the relation 
%
\begin{equation}
\label{eq:2circles}
\left[\begin{array}{r}
                             A_5\cr
                             B_5 \cr \end{array}\right]
\ = \ \frac{1}{|\beta|^2}
\left[\begin{array}{cc}
                             |\alpha|^2-\sigma\lambda &  \alpha^*\,(\lambda\sigma-1) \cr
                             \alpha\,(\lambda^*\sigma^*-1) &  |\alpha|^2 -\sigma^*\lambda^*\cr \end{array}\right]
\left[\begin{array}{r}
                             A_1\cr
                             B_1 \cr \end{array}\right]\,.
\end{equation}
}
We denote the above transition matrix as 
\begin{equation}
\label{eq:matrixT}
T 
\ = \ \frac{1}{|\beta|^2}
\left[\begin{array}{cc}
                             |\alpha|^2-\tau &  \alpha^*\,(\tau -1) \cr
                             \alpha\,(\tau^*-1) &  |\alpha|^2 -\tau^* \end{array}\right] 
= \left[\begin{array}{rrr}
                      A\phantom{^*} &  B\phantom{^*} \cr
                                      B^* &  A^* \cr \end{array}\right]\,,
\end{equation}
where $\tau=\lambda\sigma$ codes the over-all change of phase in the single ring over one turn.  
Quite interestingly, the matrix does not depend on the ratio the ring is cut by the tangency points.

\noindent 
{\bf Proposition:} The resulting matrix $T$ is an element of the  pseudo-unitary group ${\rm SU}(1,\!1)$,
the group preserving the Hermitian inner product of signature $(+1,-1)$.  That is:
\begin{equation}
\label{eq:su11}
      T^* GT = G 
     \ \hbox{where} \    
      G=\begin{bmatrix} 1&0\\0&-1 \end{bmatrix},
     \quad \det T = |A|^2 - |B|^2 = 1
\end{equation}
($T^*$ denotes the standard adjoint matrix of $N$, i.e., $N^* = \bar N^T$).
\\
\\
{\bf Proof:}  We need to show that the determinant of $T$ is 1.  
Instead of longer calculations for the resulting matrix \eqref{eq:2circles},
one may simply evaluate the product of determinant of the string of matrices \eqref{eq:1to5}:
$$
\det T = (-1)\cdot \sigma^*\lambda \cdot \frac{\beta^*}{\beta} \cdot (-1)\cdot \lambda^*\sigma \cdot \frac{\beta}{\beta^*} = 1
$$
due to $|\lambda|^2 = |\sigma|^2 = 1$ and simple cancellation.
\hfill $\square$ 
\\
\\
{\bf Corollary:}  Any power of $T$ is an element of ${\rm SU}(1,\!1)$, satisfying \eqref{eq:su11}.

~\\
Composing the total  transformation around the six rings 
(i.e., twice around the tri-ring configuration) 
is tantamount to $T^3$, which results in  
\begin{equation}
\label{eq:matrixT3}
\left[\begin{array}{r}
                             A_1\cr
                             B_1 \cr \end{array}\right]
\ = \ 
\left[\begin{array}{cc}
\!\!                            A^3 + ( 2A+A^*)\,|B|^2             &    B\, \left((A+A^*)^2 -1\right) \!\!   \cr
\!\!                             B^*\, \left((A\!+\!A^*)^2 -1\right)   &   A^{*3} + (2A^*\!+\! A)\,|B|^2     \!\!  \cr \end{array}\right]
\left[\begin{array}{r}
                           \!\!  A_1 \!\!\cr
                           \!\!  B_1 \!\! \end{array}\right],
\end{equation}
where $A$ and $B$ are defined in \eqref{eq:matrixT}.

\section{Eigenvectors as the normal modes}
\label{sec:eigenvectors}

In this section we identify the condition of resonance frequencies for the tri-microring system
by analyzing the eigenvectors of the transition matrices.
Let us start with a general mathematical property:
\\
\\
{\bf Proposition} 
A pseudo-unitary matrix 
$$
               M=  \begin{bmatrix} a &b \\ b^* & a^* \end{bmatrix} \ \in {\rm SU}(1,\!1)
$$
has an eigenvector $\mathbf v$ with eigenvalue 1 only if ${\rm Tr}\, M =2$. 
The corresponding eigenvector is (up to a complex scaling):
$$
               \mathbf v = \begin{bmatrix} b \\ 1- a \end{bmatrix} .
$$
{\bf Proof:}  Solve $\det(M-\lambda\cdot I) = 0$ for $\lambda = 1$ to get 
${\rm Re}\, a= 1$, i.e., $a=1+\varepsilon i$ for some real  $\varepsilon \in\mathbb R$.
For the second part, check by direct verification. 
\hfill $\square$   

\noindent
Let now apply it to our system.  
In case of matrix $T^3$ of Eq. \eqref{eq:matrixT3}, we get
$$
\begin{array}{rl}
          {\rm Tr}\, M & = A^3 +A^{*3} + 3(A+A^*)|B|^2 \\
                           & = (A+A^*)^3 - 3(A+A^*) \  =  \  2.
\end{array}
$$
Denote $A+A^* =x$  \ (twice the real part of $A$).
Then the above condition becomes a cubic equation
$$
                 x^3-3x-2 = 0  \, .
$$
The polynomial factors easily leading to two real solutions: $x=2,\, -1$.   
The choices for $A$ are  thus
$$
\begin{array}{rl}
\hbox{Case 1:} & A= 1+pi  \\
\hbox{Case 2:} & A =-\frac{1}{2} + pi 
\end{array}
$$
for some real $p\in \mathbb R$.
Since $T$ must be pseudo-unitary, 
\begin{equation}
\label{eq:recall}
|A|^2 - |B|^2 \ = \ 1   \, ,
\end{equation}
we may calculate both the condition for $B$ and the matrix $T$.  
\\

\noindent
{\bf Case 1:}  Substituting $A=1+pi$ to \eqref{eq:recall} gives $|B|^2=p^2$.  Hence 
$$
T=  \begin{bmatrix} 1+pi &p{\rm e}^{i\varphi} \\ p {\rm e}^{-i\varphi} & 1-pi \end{bmatrix}, 
$$where $\varphi$  is an arbitrary phase. 
Using \eqref{eq:matrixT3}, one readily gets:
$$
M = T^3 = \begin{bmatrix}              1+3pi             &  3p{\rm e}^{i\varphi} \\
                                         3p {\rm e}^{-i\varphi} & 1-3pi          \end{bmatrix} .
$$
Thus $A+A^*=2$ gives via Eq. \eqref{eq:matrixT} condition 
$$
2|\alpha|^2 - 2 |\beta|^2 = \tau + \tau^*
\qquad \Rightarrow \qquad
\tau + \tau^*  =  4|\alpha|^2 -2  
$$
and therefore the eigenvector is 
$$
\mathbf v = 
\begin{bmatrix}
2\,B  \\
A^*-A
\end{bmatrix}
\ \sim \ 
\begin{bmatrix}
2\alpha^* (\tau - 1)  \\
\tau -\tau^*
\end{bmatrix}.
$$
\\
In general the eigenvector is
$$ 
\mathbf v = 
 \begin{bmatrix}{\rm e}^{i\varphi} \\1 \end{bmatrix}. 
$$

~

\noindent
{\bf Case 2:}  Substituting $A=-1/2+pi$ to \eqref{eq:recall} gives  
$|B|^2= |A|^2 - 1 = p^2 - \frac{3}{4}$  \  
(note:  $p^2\geq 3/4$). Hence
$$
\begin{aligned}
T &=  \begin{bmatrix} -\frac{1}{2}+ip  &  \frac{\sqrt{4p^2-3}}{2}\,  {\rm e}^{i\varphi} \\ 
                                \frac{\sqrt{4p^2-3}}{2}\,  {\rm e}^{-i\varphi} & -\frac{1}{2}-ip \end{bmatrix} 
\\
&= \frac{1}{2}
\begin{bmatrix} -1+2pi  &  \sqrt{4p^2-3}\,  {\rm e}^{i\varphi} \\ 
                                \sqrt{4p^2-3}\,  {\rm e}^{-i\varphi} & -1-2pi \end{bmatrix} \,.
\end{aligned}                                
$$
Using \eqref{eq:matrixT3}, we get (somewhat expectedly):
$$
M = T^3 =        \begin{bmatrix}             1  &   0 \\ 
                                   0 & 1 \end{bmatrix}  \ 
\ \hbox{and eigenvector = anything} .
$$
Here $A+A^*=-1$ gives via Eq. \eqref{eq:matrixT}  condition 
$$
2|\alpha|^2 + |\beta|^2 = {\tau + \tau^*}
\qquad \Rightarrow \qquad
 \tau + \tau^*  =   |\alpha|^2 +1 .
$$

\section{Resonance frequencies}

The resonance frequencies are obtained by $\theta=\frac{2\pi f L }{v}$ 
where the $\theta$ denotes the phase of $\tau$. 
$$
{\rm Tr}\, T^3 = 2
\ \Rightarrow \
\left\{\!\!\!
\begin{array}{lll}
{\rm Tr}\, T = 2  &\Rightarrow \  \  \tau+\tau^* = 4|\alpha|^2-2   &  \Rightarrow \ \cos \theta = 2|\alpha|^2 -1
\\[11pt]
{\rm Tr}\, T = -1 & \Rightarrow \  \   \tau+\tau^* = |\alpha|^2+1  &   \Rightarrow \ \cos \theta = \frac{|\alpha|^2+1}{2} 
\end{array}\right.
$$

$$
\left\{\!\!\!
\begin{array}{lll}
  \Rightarrow \  \    \lambda  = \dfrac{2\pi L}{ 2k\pi \pm \arccos (2|\alpha|^2 -1)  }
& \Rightarrow \  \                f = \frac{v}{L} \,   \left( \pm \frac{\arccos (2|\alpha|^2-1)}{2\pi} + m  \right)
\\[11pt]
   \Rightarrow \  \   \lambda =  \dfrac{ 2\pi L}         {2\pi k \pm \arccos \frac{|\alpha|^2+1}{2} }
& \Rightarrow \  \               f = \frac{v}{ L} \,   \left(\pm \frac{\arccos \frac{|\alpha|^2+1}{2}}  {2\pi} + m \right)\,,
\end{array}\right.
$$
%
where $\lambda$ = permitted wavelength and $f$  = corresponding resonance frequencies.
Speed of light in the fibers = $v$.  Length of each ring = $L$ (circumference).
Also, $m\in \mathbf Z$.

\begin{figure}[ht]
\centering
\includegraphics[scale=.7]{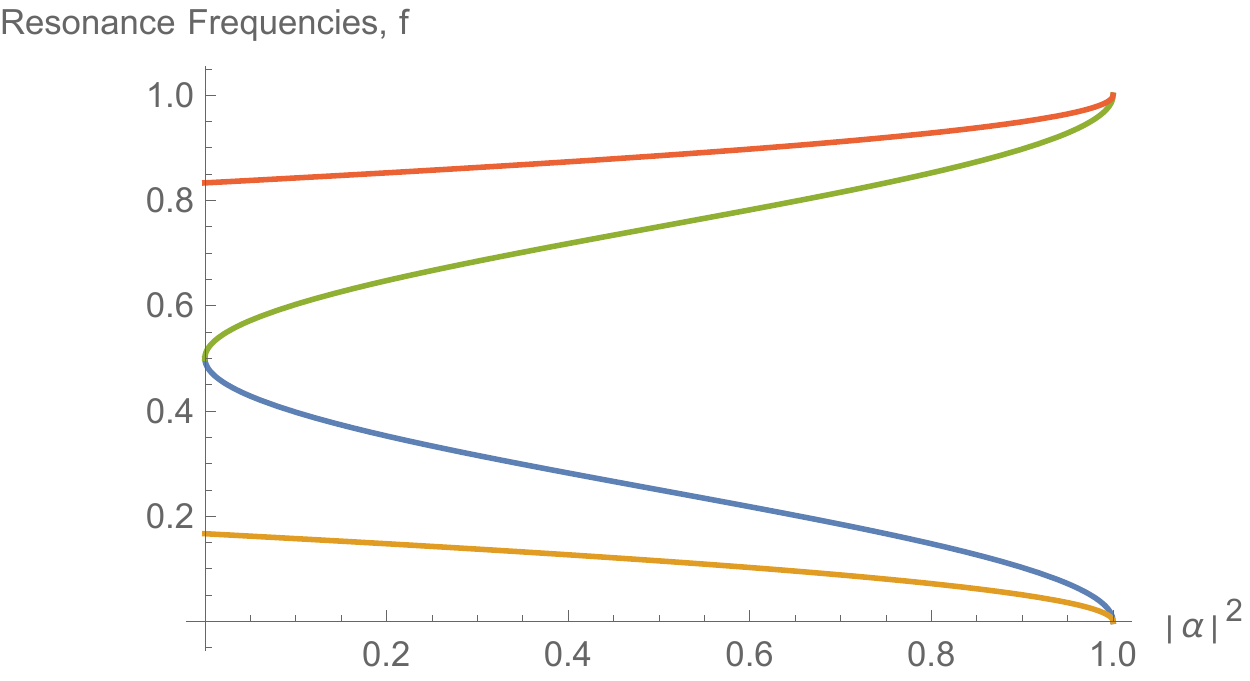}
\caption{Relation between frequency $f$ and $|\alpha|^2$.}
\label{ResFre}
\end{figure}

To illustrate this with a numerical example, assume $v/L=1$
and $|\alpha| = \sqrt{3}/2$.   
Then the series of resonance frequencies $f$ is:
$$
\begin{array}{l}
f_{1k}  =\frac{1}{6}  +m   \\[7pt]
f_{2k}  =\frac{5}{6}  + m  \\[7pt]
f_{3k}  = .08043 + m   \\[7pt]
f_{4k}  = .91957 + m .  \\
\end{array}
$$
Figure \ref{ResFre} shows the relation between the four frequencies and the power transmission coefficient. 
%

\section{Microrings with MZI couplers}

Modulating the resonator field using the waveguide-resonator coupling is known to have low-distortion, high extinction ratio, and chirp-free outputs \cite{sacher2}.  There few techniques use to modulate the coupling parameter such as coupling gap tunability \cite{Del}, dual-cavity coupled microresonator \cite{miller}, and the Mach-Zehnder interferometer (MZI) \cite{chen} techniques. Each technique has its own advantages depending on the application used.  In this section, we use the MZI technique to formulate the resonant frequency tunability characteristics of the tri-microring structure with respect to the phase variation of the MZI unit.  It is well-known that the transmission field of an MZI unit biased at quadrature changes linearly with respect to an applied filed. Another desirable feature of the MZI approach is that the output is chirp-free \cite{koyama}.

\begin{figure}[ht]
\centering
\includegraphics[scale=1]{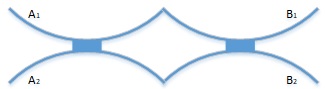}
\caption{Mach-Zehnder interferometer (MZI) structure}
\label{MZI}
\end{figure}

\begin{figure}[ht]
\centering
\includegraphics[scale=.34]{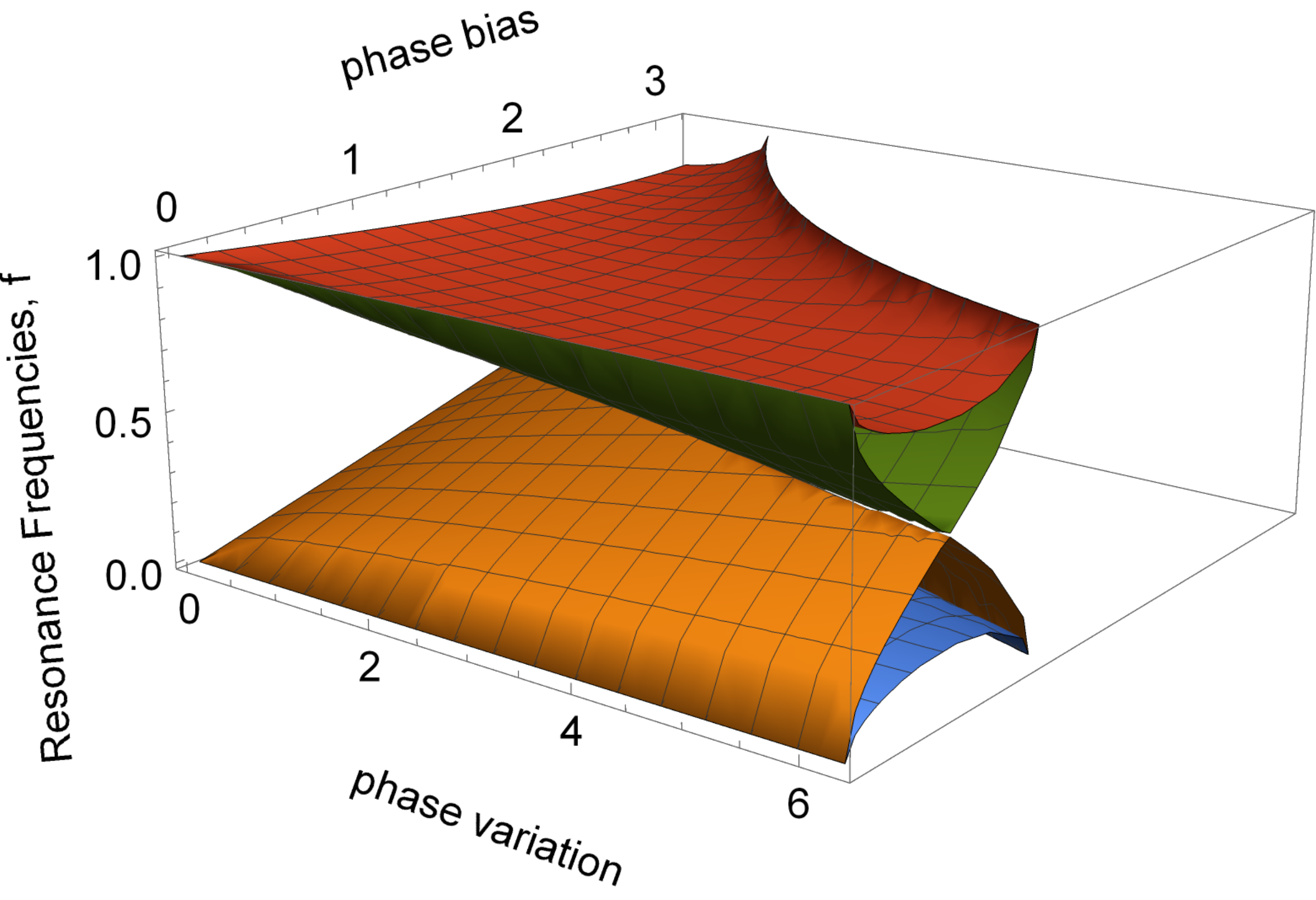}
\caption{Resonance Frequencies versus phase variation and phase bias}
\label{fre-ph0}
\end{figure}

\begin{figure}[ht]
\centering
\includegraphics[scale=.5]{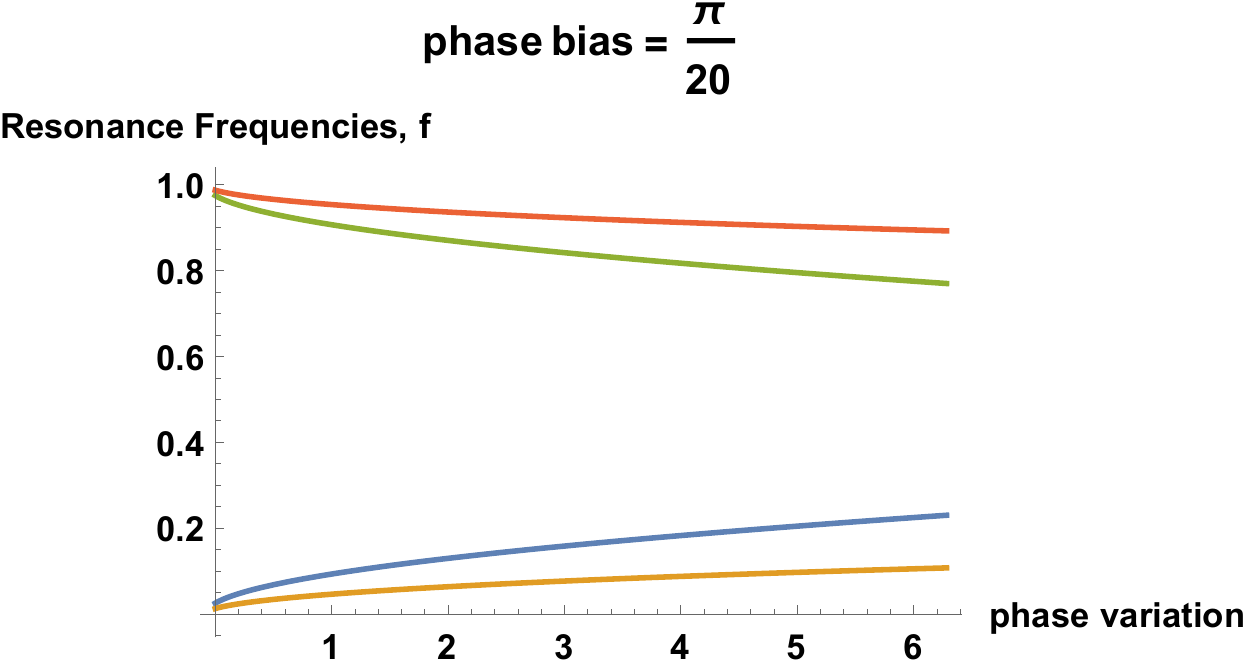}
\includegraphics[scale=.5]{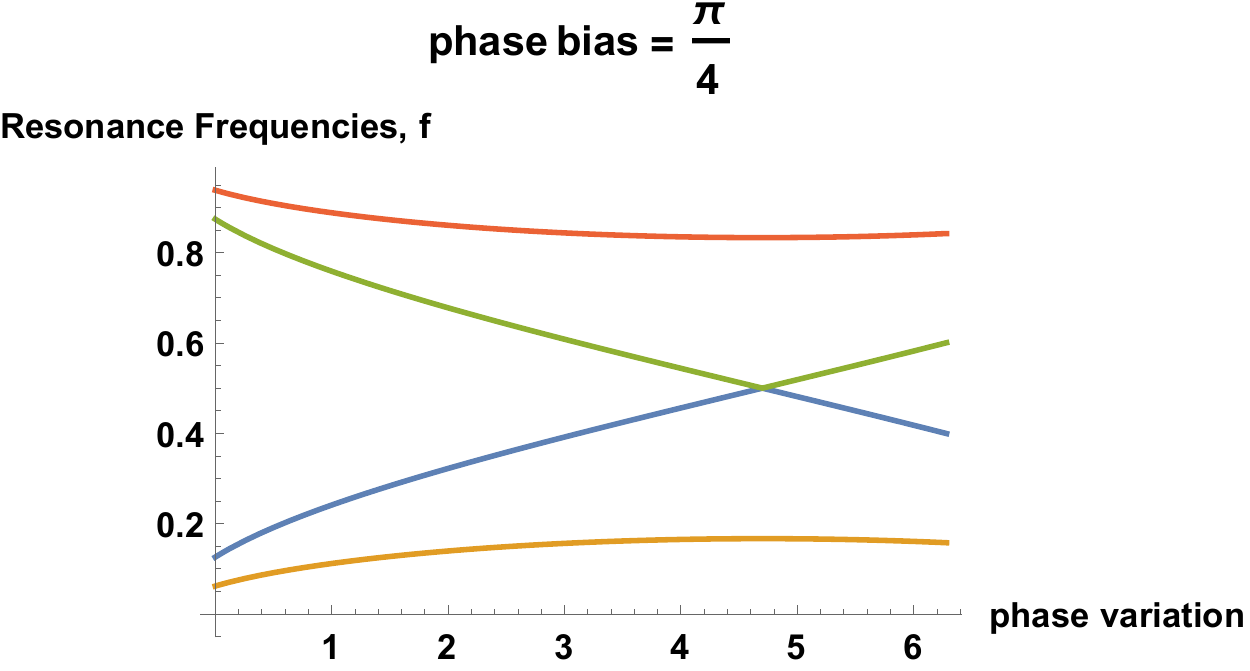}

~

\includegraphics[scale=.5]{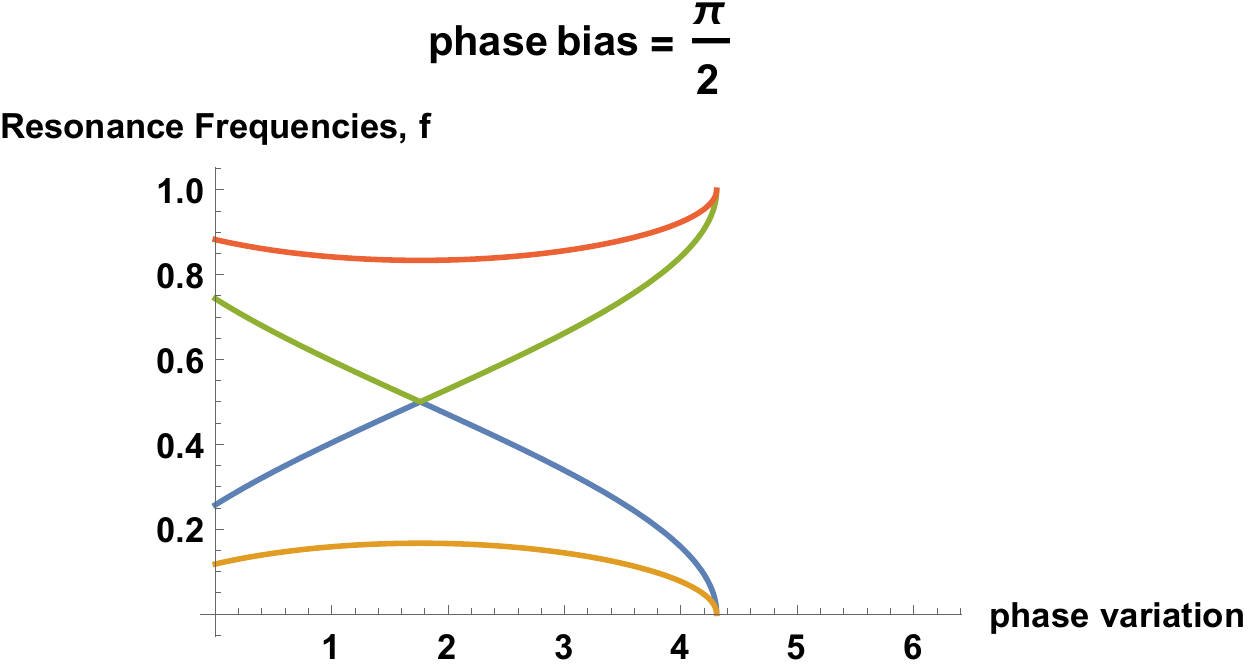}
\includegraphics[scale=.5]{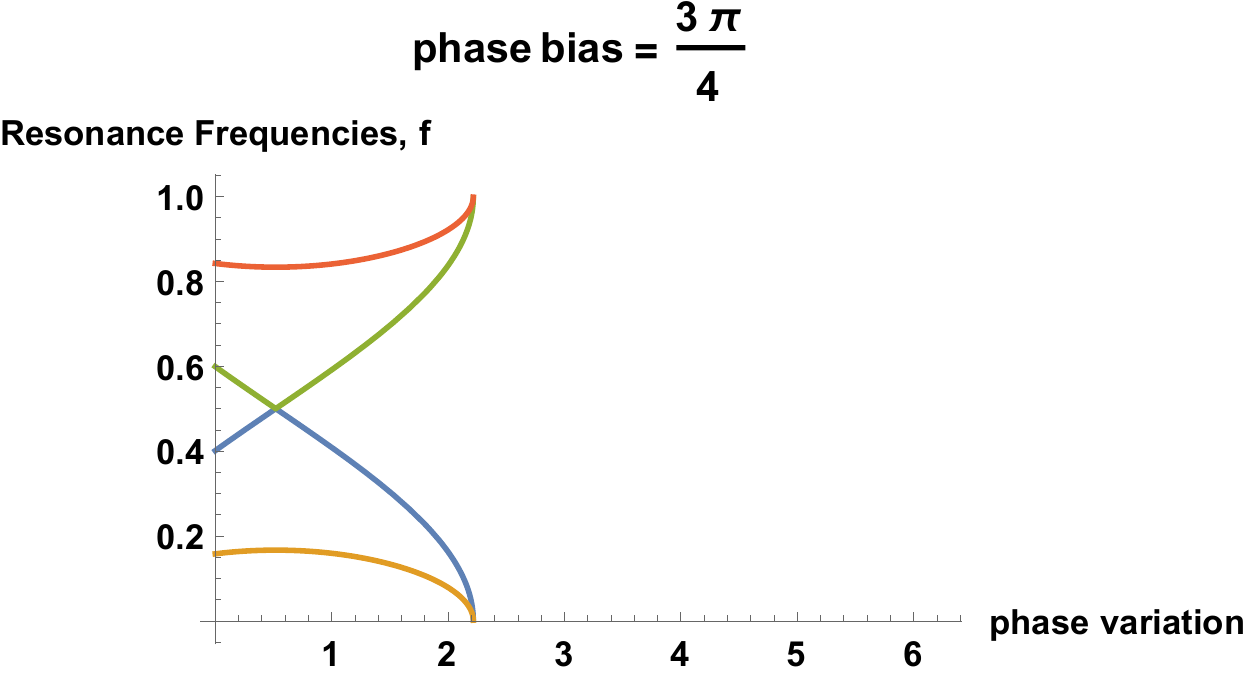}

~

\includegraphics[scale=.5]{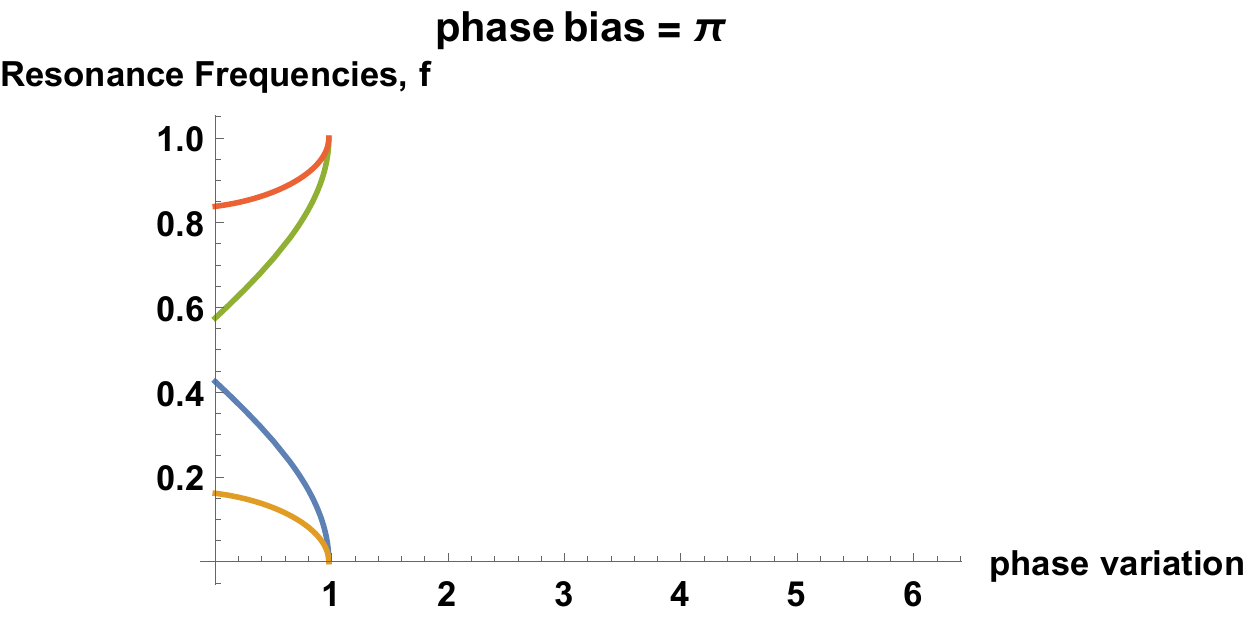}
\caption{Resonance Frequencies versus phase variation at different values of phase bias}
\label{fre-ph1}
\end{figure}

The structure of a typical MZI is shown in Fig. \ref{MZI}. There are two 3dB bi-directional couplers connected as shown.  They are biased at quadrature in the push-pull configuration, namely, the upper-leg phase is $\Delta\phi/2+\pi/2$; and the lower-leg phase is $-\Delta\phi/2-\pi/2$, where $\Delta\phi$ is the applied phase variation.  Assuming $\Delta\phi=\Delta\phi_{0}+\Delta\phi_{1}(t)$, ($\Delta\phi_{0}$ is the phase bias; and $\Delta\phi_{1}(t)$ is the phase variation.) and high-Q microrings where $\Delta\phi \ll 1$, the transmission coefficient $\alpha$ is given as \cite{sacher2}

\begin{equation}
\label{eq:phasevar}
\alpha=i(1-\frac{\Delta\phi_{0}^2}{8}-\frac{\Delta\phi_{0}\Delta\phi_{1}}{4}) \,.
\end{equation}
The resulting resonance frequencies are shown in Figure \ref{fre-ph0}. Figure \ref{fre-ph1} shows the frequency vs. the phase variation at different values of the phase bias, namely, $\pi/20$, $\pi/4$, $\pi/2$, $3\pi/4$, and $\pi$.
%
An interesting observation is that the frequencies can be controlled and tuned as close one to another as desired.   
This may have applications in the mode locking where the frequency can be modulated.  Also the tuning of the coupling via the MZI phase shift does not change the free spectral range (FSR) of the resonance resulting in a broad bandwidth modulation over the entire FSR.

\section{Summary and conclusion}

In this work, we introduced a coupled mirroring geometry in which the wavelength can be tuned by the coupling coefficient.  In practical applications, there are many simple techniques to alter the coupling coefficient such as the electro-optic modulation \cite{Coldren,Chuang}. Here the MZI approach was used for the coupling modulation. The resonance condition for the tri-microring system with identical circumferences was analyzed using the transfer matrix method. 
The resulting matrix after one complete round trip, found to be the pseudo-unitary group ${\rm SU}(1,1)$ with two eigenvectors for the unit eigenvalue.  
We examine the conditions for $\alpha$  and $\tau$  that lead to the eigenvectors as normal modes of resonance.
\\
\\
{\bf Case 1, $\tau + \tau^*  =  4|\alpha|^2 -2$ :}
The transition through from $A_1$ to $B_1$ (the first point of tangency) consists of only a change of phase.  For uncoupled-microrings case ($\alpha = 1$),  $A_1=B_1$, whereas for the complete coupling ($\alpha = 0$), $B_1=0$.  
The round-trip phase shift (RTPS) for a single ring is an even (odd) multiple of $\pi$ for uncoupled (completely coupled) scenario. 
\\
\\
{\bf Case 2, $\tau + \tau^*  = |\alpha|^2 +1$: }
The transfer matrix is the identity matrix indicating an arbitrary eigenvector provided the condition on RTPS is satisfied. 
This condition for the uncoupled case is RTPS = $2 \pi m$ whereas for the completely coupled microrings RTPS =  $2 \pi m \pm \pi /3$.

For both cases, the RTPS condition was expected for uncoupled microrings, however the result for the completely coupled ones cannot be directly obtained by following the complete round-trip path through the tri-microring geometry which results  in RTPS = $2 \pi m/3$.

Other important characteristics of resonators such as full-width-at-half-maximum (FWHM) will be investigated where $| \tau |<1$.  This work paves the ground for the Apollonian resonance in the coupled microrings with different circumferences.

\section{Appendix: Photonics and matrix groups}
\label{sec:appendix}

Coupled resonators provide an interesting application of various versions of unitary groups, known typically in other areas of physics.
The groups U(2) and  SU(2) are well known in quantum physics as the symmetry group of the two-level systems, like electron spin,
and form quantum counterpart of the rotation group of the 3D real vector space $\mathbb R^3$ \cite{Lomonaco}.
Meanwhile, their hyperbolic versions, like U(1,1) and SU(1,1), are less popular but they show up 
in hyperbolic geometry and in spinor representations of the symmetry of the Minkowski space (relativistic physics) \cite{jk2}.
It is thus interesting coupled resonators imitate involve these groups and moreover provide 
an intriguing example of transition from regular to hyperbolic version via reconsideration of the analysis of the couplings: 
from in-out to waveguide-to-waveguide.

What is even more pleasing, the coupling resonators  hint at extension of these groups, denoted here by stars,  
${\rm SU}^*$ and ${\rm U}^*$.  
These groups are introduced in this paper to describe the waveguide-to-waveguide transitions 
in order to use the group-theoretic properties to facilitate calculations.
Below, we give a short review of these objects.
\\

\noindent
{\bf A. Unitary and special unitary groups.}  
A (finite dimensional) Hilbert space $\mathcal H$  is a complex linear space equipped with a sesquilinear inner product, 
i.e., a product that has complex values and satisfies
\begin{equation}
\label{a:hermitian}
\langle v,\, w\rangle =  \overline{ \langle w,\, v\rangle }
\end{equation}
(the bar denotes complex conjugation).
In the presence of basis, vectors are matrix columns and the product has a form
\begin{equation}
\label{a:vw}
\langle v,\,w\rangle  = v^* G w
\end{equation}
for some square matrix $G$.  
One may choose a basis in which matrix $G$ is a ``unit matrix'', for instance in 2 dimension:
\begin{equation}
\label{a:G}
G = \left[\begin{array}{cr}
                             1  & 0   \\
                             0  &  1        \end{array}\right] \,,
\end{equation}
and therefore becomes ``invisible'' in the defining equation \eqref{a:vw}:
\begin{equation}
\label{a:viwi}
\langle v,\,w\rangle  = v^* w  = \bar v_1\bar w_1 +  \bar v_2\bar w_2 \,.
\end{equation}

Unitary group is the group of linear transformations of the space $\mathcal H$ that preserve
the inner product
\begin{equation}
\label{a:MvMw}
\langle Mv,\, Mw\rangle = \langle v,\, w\rangle \,.
\end{equation} 
From this, in matrix notation:
\begin{equation}
\label{a:MGM}
M^*GM = G
\end{equation}
and again, in appropriate basis $G$ becomes invisible:
\begin{equation}
\label{a:MM}
                             M^* = M^{-1} \,.
\end{equation}
This implies that the determinant of this matrix, as a complex number satisfies $|\det M| = 1$.
We may thus reduce the group by adding the condition that the determinant is one.  This leads to the 
special unitary group.  In matrix representation:
\begin{equation}
\label{a:MMdet}
                              M^* = M^{-1},  \qquad \det M= 1 \,.
\end{equation}
In particular, for the 2 dimensional Hilbert space in the standard basis in which $G$ looks like identity matrix, 
the typical element is 
\begin{equation}
\label{a:M}
                   M = \left[\begin{array}{rrr}
                             \alpha\phantom{^*} &  \beta\phantom{^*} \cr
                             -\beta^* &  \alpha^* \cr \end{array}\right]
\end{equation}
is an element of the special unitary group ${\rm SU}(2)$.  In particular,
\begin{equation}
\label{a:MMI}
           M^*M= I
           \quad\hbox{and}\quad
         \det M = |\alpha|^2 + |\beta|^2 = 1 \,,
\end{equation}
where $M^*$ denotes Hermitian conjugation, that is $M^* = \bar M^T$  \ 
(transposition and complex conjugation of the entries).
\\

\noindent
{\bf B.  Pseudo-unitary groups.}
Other versions of the unitary groups arise when one replaces the inner product by a more general form.
For instance, in two dimensional Hilbert space, matrix  
\begin{equation}
\label{a:Gx}
                       G = \left[\begin{array}{cr}
                             1  & 0   \\
                             0  &  -1        \end{array}\right]
\end{equation}
defines a hyperbolic inner product and corresponding group called ``{\bf pseudo-unitary group}'' via \eqref{a:MvMw}.
The matrix equation \eqref{a:MGM} still holds and still implies $\det M=1$.
Also as before, the {\bf special unitary group} SU(1,1) is obtained by 
restriction of U(2) by demanding that the determinant is equal to one.
A typical element of $SU(1,1)$ is 
\begin{equation}
\label{a:Mx}
                             \left[\begin{array}{cc}
                             \alpha &\beta    \\
                             \beta^* & \alpha^*      \end{array}\right] \,.
\end{equation}

~

\noindent
{\bf C.  Extended pseudo-unitary groups.}
Motivated by the model analyzed in the present paper, 
wet we consider an extended versions the pseudo-unitary groups by relaxing the defining property to 
\begin{equation}
\label{a:N}
                                   N^*\,G\,N \ = \ \pm G\,.
\end{equation}
Note $\det G = \det(-G)$, hence the above determines as in the usual unitary case that $|\det N | = 1$.
Thus we have two cases:
Extended pseudo-unitary group and extended special pseudo-unitary group;
in the 2-dimensional case
\begin{equation}
\label{a:U11}
                               U^*(1,1) \quad  \hbox{and} \quad SU^*(1,1) \,.
\end{equation}
Topologically, these groups are generated by inclusion of an additional element that flips the positive and negative 
directions.  For instance, if  
\begin{equation}
\label{a:F}
                           F = \left[\begin{array}{cr}
                             0  & 1   \\
                             1  & 0        \end{array}\right]
\end{equation}
then the extended versions of the groups may be viewed as generated by $F$ and the standard element:
\begin{equation}
\label{a:U111}
\begin{array}{cl}
           U^*(1,1) \ &= \ {\rm gen}\/ \{\,F, g\in U(1,1) \,\}  \\
          SU^*(1,1) \ &= \ {\rm gen}\/ \{\,F, g\in SU(1,1) \,\} \,.
\end{array}
\end{equation}

Below, the diagram represents mutual dependence of these groups.
The map from matrix \eqref{eq:inout} to \eqref{eq:transN} in Sec. \ref{sec:basics},
namely $m: SU(2) \ \to \ U^*(1,1)$:
\begin{equation}
\label{a:SUU}
        \left[\begin{array}{rrr}
                             \alpha\phantom{^*} &  \beta\phantom{^*} \cr
                             -\beta^* &  \alpha^* \cr \end{array}\right]
        \quad\mapsto \quad
        \frac{1}{\beta}
        \left[\begin{array}{cc}
                             -\alpha &1   \\
                             -1 & \alpha^*      \end{array}\right]  \,,
\end{equation}
is well defined only for the elements with $\beta \not=0$ and is not a homomorphism.
%
\begin{equation}
\begin{tikzpicture}[baseline=-0.8ex]
    \matrix (m) [ matrix of math nodes,
                         row sep=2em,
                         column sep=4em,
                         text height=3.8ex, text depth=3ex] 
   {
   \displaystyle \ {\rm SU(2)} \quad   
                    & \quad {\rm SU^*(1,1)}  \quad  
                             & \quad   {\rm SU(1,1)}                            \\
    \displaystyle \ {\rm U(2)} \quad
                    & \quad {\rm U^*(1,1)} \quad
                             & \quad {\rm U(1,1)}      \\
     };
  \path[->]
        (m-1-3) edge node[above] {$\subset$} (m-1-2)
        (m-2-3) edge node[above] {$\subset$} (m-2-2)
        (m-1-1) edge node[right]  {$\subset$} (m-2-1)
        (m-1-2) edge node[right] {$\subset$} (m-2-2)
        (m-1-3) edge node[right] {$\subset$} (m-2-3);
    \path[o->]
        (m-1-1) edge node[right] {$m$} (m-2-2)
;

\end{tikzpicture}   
\end{equation}

\noindent
The appearance of the discussed groups in coupled waveguides and microrings are indicated below.  
\\
\\
1. The matrices defining the in-out signal transition belong to group ${\rm  SU}(2)$.
\\\\
2.  The matrices representing the waveguide-to-waveguide transitions are elements of  ${\rm U}^*(1,1)$.
The typical element  
\begin{equation}
\label{eq:transxMM}  
N = 
\frac{1}{\beta}
\left[\begin{array}{cc}
          -\alpha & 1   \\
               -1 & \alpha^*      \end{array}\right]
\end{equation}
satisfies $N^*\,G\,N \ = \ -G$:
\begin{equation}
\label{eq:su11x}
\frac{1}{\beta^*}
\begingroup 
\setlength\arraycolsep{2.5pt}
\left[\begin{array}{cr}
                             -\alpha^*  & -1   \\
                              1              & \alpha        \end{array}\right]
\left[\begin{array}{cr}
                             1  & 0   \\
                             0  &  -1        \end{array}\right]
\;\frac{1}{\beta}
\left[\begin{array}{cc}
                             -\alpha &1   \\
                             -1 & \alpha^*      \end{array}\right]
 = 
\left[\begin{array}{cr}
                             -1  & 0   \\
                             ~~0  & 1        \end{array}\right]
\endgroup
\end{equation}
with the determinant being $\det N = \beta^*/\beta \   \in S^1 \subset \mathbb C$.
%
%

~\\
3. Propagation along a waveguide contributes phase change by some complex factor 
$\sigma\in \mathbb C$,  \, $|\sigma|^2 = 1$. Two waveguides (of possibly different lengths and optical properties), correspond to a diagonal matrix:
\begin{equation}
\label{eq:phase00}
\left[\begin{array}{cc}
                             \sigma &  0 \cr
                             0 &  \lambda \cr \end{array}\right]
\,,  \qquad |\sigma|^2 = |\lambda|^2 = 1 \,. 
\end{equation}
As such it belongs to both ${\rm SU}(2)$ and ${\rm SU}^*(1,1)$.

Here is the summary of the matrix characterization of the various unitary groups: 

$$
\begin{array}{lll}
{\rm U}(2):         &\quad  M^*M=I &\quad   \Rightarrow |\det M| =1\\
{\rm SU}(2):       &\quad  M^*M=I  &\quad \hbox{and} \ \det M=1\\[7pt]
{\rm U}(1,1):      &\quad  M^*GM=G &\quad   \Rightarrow  |\det M|=1\\
{\rm SU}^*(1,1)   &\quad  M^*GM=G & \quad \hbox{and} \  \det M=1\\[7pt]
{\rm U}^*(1,1):   &\quad  M^*GM= \pm G &\quad   \Rightarrow  |\det M|=1\\
{\rm SU}^*(1,1): &\quad M^*GM= \pm G  &\quad \hbox{and} \  \det M=1
\end{array}
$$
Clearly, the group properties assure that the products of the group elements 
preserve the defining group properties.  We used this fact in Section \ref{sec:3circles}.


\end{document}